\documentclass[prl,twocolumn,showpacs,amsmath,aps]{revtex4}
\usepackage{graphicx}  % needed for figures
\usepackage{bm}        % for math
\usepackage{amssymb}   % for math
\usepackage{amsmath}
\usepackage{enumerate}
\usepackage{color}
\usepackage{pgfplots}

\begin{document}

\title{Heat fluctuations for underdamped Langevin dynamics}
\author{ M.L. Rosinberg}
\affiliation{Laboratoire de Physique Th\'eorique de la Mati\`ere Condens\'ee, Universit\'e Pierre et Marie Curie, CNRS UMR 7600,\\ 4 place Jussieu, 75252 Paris Cedex 05, France}
\email{mlr@lptmc.jussieu.fr}
\author{G. Tarjus}
\affiliation{Laboratoire de Physique Th\'eorique de la Mati\`ere Condens\'ee, Universit\'e Pierre et Marie Curie, CNRS UMR 7600,\\ 4 place Jussieu, 75252 Paris Cedex 05, France}
%\email{tarjus@lptmc.jussieu.fr}
\author{T. Munakata}
\affiliation{Department of Applied Mathematics and Physics, 
Graduate School of Informatics, Kyoto University, Kyoto 606-8501, Japan} 
%\email{tmmm3rtk@hb.tp1.jp}

\begin{abstract}
 Fluctuation theorems play a central role in nonequilibrium physics and stochastic thermodynamics. Here we derive an integral fluctuation theorem for the dissipated heat in systems governed by an underdamped Langevin dynamics. We show that this identity may be used to predict the occurrence of extreme events leading to exponential tails in the probability distribution functions of the heat and related quantities.  
 \end{abstract} 


\maketitle

\section{Introduction}

The discovery of fluctuation theorems (FT's) has considerably improved our understanding of nonequilibrium physics by revealing the universal properties of the probability distribution functions (pdf's) for thermodynamics quantities such as heat, work, or entropy  measured over a time interval $t$ (see \cite{S2012} for a review and references therein). In short, a  FT  states that positive fluctuations of an observable  exponentially  dominate negative ones, which is  experimentally observed in small stochastic systems, {\it e.g.}, a dragged Brownian particle or a noisy electrical circuit. 

Physical observables with identical expectation values do not always obey the same FT, as was first pointed out in \cite{VC2003} for an overdamped particle in a moving harmonic trap. In this case, all fluctuations of the work ${\cal W}$ in the nonequilibrium stationary state (NESS) satisfy the so-called  ``conventional" FT in the long-time limit ({\it i.e.}, the large deviation function satisfies the Gallavotti-Cohen symmetry \cite{GC1995,LS1999,K1998}), but this is not true for the dissipated heat ${\cal Q}$ due to rare but  large fluctuations in the internal energy difference $\Delta {\cal U}$.  This gives rise to exponential tails in the pdf of ${\cal Q}$ and is signaled by the presence of  singularities in the corresponding characteristic function. Such temporal ``boundary" effects  that take place in systems with unbounded potentials are now well-documented, both theoretically \cite{Fa2002,V2006,BJMS2006,PRV2006,HRS2006,TC2007,RH2008,N2012,NP2012,LKP2013,N2014,KKP2014} and experimentally \cite{GC2005,JGC2007,CJP2010}.

The occurrence of extreme events in heat fluctuations, associated with exponential tails in the pdf, is a significant feature that may be relevant to the functioning of small devices, for instance electrical nanocircuits \cite{VCC2004}. However,  there is yet no general principle that states when such tails exist. In this Letter, we take a step in this direction by deriving an integral fluctuation theorem (IFT) for the dissipated heat in Langevin systems that has been overlooked in the  stochastic thermodynamics literature so far. This IFT takes a simple and universal form for an underdamped motion and holds for any observation time and any initial condition.  A similar identity exists in the overdamped limit, but only for linear dynamics. We show that this IFT, by imposing a constraint on the pdf of ${\cal Q}$, can be used in a NESS and in the long-time limit to predict (at least partially) the existence of exponential tails. This is illustrated  by analytical calculations for a harmonic chain connected to reservoirs at different temperatures and for a Brownian particle subjected to a non-Markovian feedback control.

\section{Integral fluctuation theorem}

 We consider an ensemble of $N$  particles with mass $m_i$ ($i=1,\cdots,N$) in $d$ dimensions, each one being coupled to a heat reservoir in equilibrium at inverse temperature $\beta_{i}=1/T_{i}$ (Boltzmann's constant is set to unity throughout the Letter). The dynamics is described by the set of $N$ coupled equations
 \begin{align} 
 \label{EqL}
m_{i}\dot{ \bf v}_{i}={\bf F}_{i}([{\bf r}],t)-\gamma_{i} {\bf v}_{i} +{\boldsymbol\xi}_{i}\ ,
\end{align}
where  ${ \bf v}_{i}=\dot{ \bf r}_{i}$, ${\bf r}=({\bf r}_1,{\bf r}_2,\cdots,{\bf r}_N)$, ${\bf F}_{i}([{\bf r}],t)$ is the total force acting on particle $i$, $\gamma_i$ its friction coefficient, and the ${\boldsymbol\xi}_{i}$'s are Gaussian white noises with zero mean and variances $\langle \xi_{i\mu}(t)\xi_{j\nu}(t')\rangle=2D_i \delta_{\mu \nu}\delta_{ij}\delta (t-t')$ with $D_i=\gamma_i T_i$ and $\mu,\nu=1,\cdots,d$.   The notation ${\bf F}_{i}([{\bf r}],t)$ indicates that the force at time $t$ may depend on the previous positions of the particles, for instance at a time $t-\tau$ where $\tau$ is a delay.

Let ${\bf X}$ denote a trajectory of the system in phase space  that starts at the point ${\bf x}^0=({\bf r}, {\bf v})_{t=0}$ and is observed during the time interval $[0,t]$. (For ease of notation, we derive the IFT for a Markovian evolution, but it is straightforwardly generalized to the non-Markovian case with delay.)   The conditional probability of ${\bf X}$ is  given by
\begin{align}
\label{Eqpath0}
{\cal P}[{\bf  X}\vert  {\bf x}^0]\propto \prod_{j} e^{\frac{d\gamma_j}{2m_j}t} e^{-\beta_j {\cal S}_j[{\bf X}]} \ ,
\end{align}
where
\begin{align}
\label{Eqaction1}
\beta_j {\cal S}_j[{\bf  X}]=\frac{1}{4D_j}\int_{0}^{t}dt' \:\Big[m_j\dot {\bf v}_j+\gamma_j {\bf v}_j-{\bf F}_j([{\bf r}],t')\Big]^2 
\end{align}
is an Onsager-Machlup action functional \cite{OM1953} and the exponential factors $e^{\frac{d\gamma_j}{2m_j}t}$ come from the Jacobian of the transformations ${\boldsymbol\xi}_{j}(t) \rightarrow {\bf r}_{j}(t)$ in the continuum limit (see \cite{IP2006} and the Supplemental Material \cite{SM}). Singling out these factors is crucial for the forthcoming argument. We recall that there is no need to specify the convention for the stochastic calculus as long as $m_i\ne 0$ for all Brownian particles. 

Following \cite{S1998}, we define the heat dissipated into the bath $i$ during the time interval $[0,t]$ by the functional 
\begin{align}
\label{EqQ}
{\cal Q}_i[{\bf X}]=\int_0^t dt'\: \Big[\gamma_i {\bf v}_i-{\boldsymbol\xi}_i]  {\bf v}_i 
=\int_0^t dt'\:\Big[-m_i \dot {\bf v}_i+{\bf F}_i]  {\bf v}_i
\end{align}
where scalar products are implicit and the integral is interpreted with the Stratonovich prescription. Like for an overdamped motion, ${\cal Q}_i[{\bf X}]$ identifies with the logratio between the probability of ${\bf X}$  and that of its time-reversed image ${\bf X}^\dag$, conditioned on their initial points \cite{IP2006}. Such local detailed balance equation is at the core of  all FT's based on time reversal. The elementary observation that motivates the present Letter is that ${\cal Q}_i[{\bf X}]$ can be also expressed as a logratio of path probabilities {\it without} referring to time reversal. Instead, one considers an auxiliary dynamics in which the friction coefficient $\gamma_i$ for particle $i$ is changed into $-\gamma_i$ while keeping    $D_i$ fixed. The conditional  probability of a trajectory ${\bf X}$ is then expressed as
\begin{align}
\label{Eqpath1}
\hat{\cal P}[ {\bf X}\vert {\bf x}^0]\propto e^{-\frac{d\gamma_i}{2m_i}t} e^{-\beta_i\hat{\cal S}_i[{\bf X}]} \prod_{j\ne i} e^{\frac{d\gamma_j}{2m_j}t} e^{-\beta_j {\cal S}_j[{\bf X}]} \ ,
\end{align}
where $\beta_i\hat{\cal S}_i[{\bf  X}]=\beta_i{\cal S}_i[{\bf  X}]_{\gamma_i \rightarrow -\gamma_i}$ 
and the minus sign in the argument of the first exponential factor results from the change $\gamma_i\to -\gamma_i$ \cite{SM}.  (From now on, the hat symbol  will  refer to this auxiliary dynamics.) Comparing with Eq. (\ref{Eqpath0})  immediately  leads to the relation
\begin{align}
\label{Eqratio}
\frac{{\cal P}[{\bf X}\vert {\bf x}^0]}{ \hat {\cal P}[ {\bf X}\vert {\bf x}^0]}=e^{\frac{d\gamma_i}{m_i}t}e^{\beta_i  {\cal Q}_i[{\bf X}]} \ .
\end{align}
This can also be viewed as an application of the Girsanov formula, which however requires a careful treatment of the continuum limit of the path probabilities\cite{SM}. This relation in turn implies the integral fluctuation theorem (IFT)
\begin{align}
\label{EqIFTQ}
\langle e^{-\beta_i{\cal Q}_i[{\bf X}]}\rangle_0&=\int {\cal D}{\bf X} \:  e^{-\beta_i  {\cal Q}_i[{\bf X}] } {\cal P}[ {\bf X}\vert {\bf x}^0]=e^{\frac{d\gamma_i}{m_i}t}
\end{align}
where  $\langle...\rangle_0$ denotes an average over all possible paths ${\bf X}$ with arbitrary initial point ${\bf x}^0$. 
This nonequilibrium identity is the main result of this Letter.

It is clear that changing the sign of several $\gamma_i$'s together leads to other remarkable identities such as $\langle e^{-(\beta_i{\cal Q}_i+\beta_j {\cal Q}_j)}\rangle_0= e^{d(\frac{\gamma_i}{m_i}+\frac{\gamma_j}{m_j})t}$, etc... Note also that the argument can be easily generalized to cases where a particle is in contact with several baths  and/or several particles are in contact with the same bath.

From Jensen's inequality, the IFT implies that $\beta_i \langle {\cal Q}_i\rangle_0\ge -d (\gamma_i/m_i)t$,  which is actually a trivial inequality that can be recovered by taking the average of Eq. (\ref{EqQ}) over the noises history, $\langle {\cal Q}_i\rangle_0 =d(\gamma_i/m_i) \int_0^t dt'\: (T_i^{(v)}(t')-T_i)$, with $T_i^{(v)}(t')=m_i\langle  {\bf v}_i^2(t)\rangle_0>0$. As usual, the IFT is more informative since it implies  that there are trajectories for which $\beta_i {\cal Q}_i<-d (\gamma_i/m_i)t$. 

For an overdamped motion, the counterpart of the sign reversal of $\gamma_i$ is the sign reversal of the mobility ${\cal M}_i$ (which is assumed to be isotropic for simplicity). This leads to a simple result for linear forces only and the IFT then reads $\langle e^{-\beta{\cal Q}_i}\rangle_0=e^{{\cal M}_i \epsilon_i t}$ where $\epsilon_i=-\sum_{\mu=1}^d \partial F_{i\mu}/\partial r_{i\mu}$. One can check that this identity is verified by all linear diffusion systems studied so far, both theoretically \cite{VC2003,V2006,BJMS2006,NP2012,LKP2013,N2014,KKP2014} and experimentally \cite{GC2005,CINT2013}, although it was unnoticed \cite{BF2015}.An example is given in  the Supplemental Material\cite{SM}.

\section{Application to the prediction of rare events}

We now show that the IFT [Eq. (\ref{EqIFTQ})] has interesting consequences  for the pdf's of the stochastic heat and related quantities. We assume that the system has reached  a NESS and focus on the long-time limit.

Let $P_A(A)=\langle \delta({\cal A}[{\bf X}]-A)\rangle$ denote the  pdf of the fluctuating quantity $\mathcal A$, {\it e.g.}, the heat ${\cal Q}_i$,  where the average is over all possible trajectories with an initial state drawn from a distribution $p({\bf x}^0)$ (later taken as $p_{st}$). In the long-time limit, this pdf is expected to satisfy a large deviation principle \cite{T2009}, $P_A(A=at) \sim  e^{-I_A(a)t+o(t)}$, where $I_A(a)$ is the large-deviation function (LDF). Similarly, the  characteristic (or moment generating)  function $Z_A(\lambda,t)\equiv \langle e^{-\lambda {\cal A}[{\bf X}]}\rangle_{st}=\int_{-\infty}^{\infty} dA \: e^{-\lambda A} P_A(A)$  behaves asymptotically as $Z_A(\lambda,t)\sim e^{\mu_A(\lambda)t}$ where $\mu_A(\lambda)$ is the scaled cumulant generating function (SCGF) given by the largest eigenvalue of the appropriate Fokker-Planck operator. At first sight, $\mu_A(\lambda)$ should be the same function for the whole class  $[{\cal A}]_{{\cal Q}_i}$ of observables that differ from {${\cal Q}_i$} by only temporal boundary terms. However, \cite{VC2003} and subsequent work\cite{Fa2002,V2006,BJMS2006,PRV2006,HRS2006,TC2007,RH2008,N2012,NP2012,LKP2013,N2014,KKP2014} showed that this is not always true. To see this, it is convenient to rewrite the characteristic function for ${\cal A}\in [{\cal A}]_{{\cal Q}_i}$ under the alternative form 
\begin{align}
\label{EqZA}
Z_A(\lambda,t)\sim g_A(\lambda) e^{\mu(\lambda)t}\ ,
\end{align}
where $\mu(\lambda)$ is computed by neglecting the influence of the initial and final states, for instance by imposing periodic boundary conditions. In the case of linear systems, one can then solve the equations of motion by Fourier transform, as done in the two examples discussed below. In consequence, $\mu(\lambda)$ in Eq. (\ref{EqZA}) does not depend on the observable ${\cal A}$, and the same is true for the LDF $I(a)$ defined by the Legendre transform $I(a)=-[\mu(\lambda^*(a))+\lambda^*(a)a]$, with the saddle point $\lambda^*$ determined by $\mu'(\lambda^*(a))=-a$\cite{T2009}. On the other hand, the pre-exponential factor in Eq. (\ref{EqZA}) does depend on ${\cal A}$ since it  results from an average over the initial and final states (see below for a refined analysis).  The important point it that this function [specifically $g_Q(\lambda)$] may diverge for certain values of $\lambda$ in the region of the saddle-point integration. The actual SCGF $\mu_A$ then differs from $\mu$ for these  values of $\lambda$ and the actual LDF $I_A$ differs from $I$. We now show that this information can be deduced from the IFT [Eq. (\ref{EqIFTQ})], at least for $\lambda=\beta_i$, with no need to investigate the analytical properties of  the pre-exponential factor $g_A(\lambda)$.

To avoid an overly formal discussion, we  first consider two specific cases, which correspond respectively to a Markovian dynamics and to a non-Markovian one with delay. A more general statement will be given in the conclusion. 

\subsection{Heat flow in harmonic chains}  

Our first example is a harmonic chain connected at its two ends to reservoirs at different temperatures $T_L$ and $T_R$.  This is a simple model for heat conduction in which fluctuations can be exactly computed in the long-time limit \cite{KSD2011,SD2011,FI2012,Sab2012}. This amounts to setting all  $\gamma_i$'s and $T_i$'s to zero in Eq. (\ref{EqL}) except $\gamma_1=\gamma_L,\gamma_N=\gamma_R$, $T_1=T_L,T_N=T_R$ (accordingly, the products over $j$  in Eqs. (\ref{Eqpath0}) and (\ref{Eqpath1}) are also restricted to $j=1,N$).  For simplicity, we take all particles with the same mass $m$ and focus on the one-dimensional case but this can be extended to different masses and $d$ dimensions. 
The system of $N$ coupled Langevin equations reads:
\begin{align}
\label{eq_SI_langevin}
&m \dot{v}_{1}=k(u_2-2u_1)-\gamma_L v_{1} + \xi_{1}\nonumber\\
&m \dot{v}_{i}=k(u_{i+1}+ u_{i-1}-2u_i) \,,\;\;\; i=2,\cdots,N-1\nonumber\\
&m \dot{v}_{N}=k(u_{N-1}-2u_N)-\gamma_R v_{N} + \xi_{N} \ ,
\end{align}
 where $u_i$ is the displacement about the equilibrium position and $k$ is the spring constant. In the NESS, the main quantity of interest is the heat exchanged between the system and one of the reservoirs, say ${\cal Q}[{\bf X}]\equiv{\cal Q}_L[{\bf X}]=\int_0^t dt'\:[\gamma_1v_1(t')-\xi_1(t')] v_1(t')$. We may also consider the medium entropy production $\Sigma_m[{\bf X}]=\beta_L {\cal Q}_L[{\bf X}]+\beta_R {\cal Q}_R[{\bf X}]$ and the total entropy production $\Sigma[{\bf X}]=\Sigma_m[{\bf X}]+\ln [p_{st}({\bf x}^0)/p_{st}({\bf x}^t)]$ where $p_{st}({\bf x})$ is the nonequilibrium stationary pdf.

A direct consequence of Eq. (\ref{EqIFTQ}) is that the SCGF $\mu_Q(\lambda)$ is equal to $\gamma_L/m$ for $\lambda=\beta_L$. Therefore, one could naively think that $\mu(\beta_L)$ in Eq. (\ref{EqZA}) is also equal to $\gamma_L/m$. However, this is not always true. To see this, we start from the analytical expression of $\mu(\lambda)$ computed in Ref. \cite{KSD2011}: $\mu(\lambda)=-(1/4\pi)\int d\omega \:\ln [1+{\cal T}(\omega)T_LT_R \lambda(\beta_L-\beta_R-\lambda)]$ where  ${\cal T}(\omega)$ is the  phonon transmission function whose expression is recalled in the Supplemental Material \cite{SM}. For the special value $\lambda=\beta_L$, it is then easy to show that  \cite{SM}
 \begin{align}
\label{Eqmu1}
\mu(\beta_L)=\int \frac{d\omega}{2\pi}\ln \frac{\vert \det \boldsymbol{\hat \chi}(\omega)\vert}{\vert \det \boldsymbol \chi(\omega)\vert}
\end{align}
where $\boldsymbol \chi(\omega)$ is the matrix formed by the response functions of the $N$ harmonic oscillators and $\boldsymbol{\hat \chi}(\omega)$ is formally defined by  changing the sign of $\gamma_L$ in $\boldsymbol \chi(\omega)$.  
 \begin{figure}[hbt]
\begin{center}
\includegraphics[width=8cm]{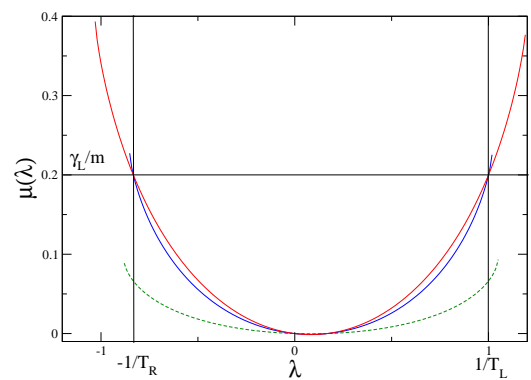}
 \caption{\label{Figmulambda} (Color on line) The function $\mu(\lambda)$  for a chain of $N=3$ harmonic oscillators with $\gamma_L=0.2,\gamma_R=1$, $T_L=1,T_R=1.2$ [$\mu(\lambda)$ is defined in an interval $(\lambda_-,\lambda_+)$ with diverging slopes at the boundaries]. Each curve corresponds to a different value of the spring constant: $k=5$ (red), $1$ (blue), $0.1$ (green dashed). For $k\geq k_c=0.6$, one has $\mu(\beta_L)=\gamma_L/m$.}
\end{center}
\end{figure}

Two cases may happen as illustrated in Fig. 1 where we plot $\mu(\lambda)$ for $N=3$. First, the hat dynamics defined by the change $\gamma_L\to -\gamma_L$ ``converges", {\it i.e.}, the system reaches a steady state independent of initial conditions. A necessary condition is that $\gamma_R>\gamma_L$, as shown rigorously in \cite{SM}. (Intuitively, $\gamma_R$ must  be large enough to damp the instability in the dynamics created by the change $\gamma_L\to -\gamma_L$.) However, this is not sufficient and the coupling $k$ must also be larger than a critical value $k_c$ (equal to $0.6$ in the example of Fig. 1). The convergence of the hat dynamics means that the response functions of the $N$ oscillators in the time-domain, defined as the inverse Fourier transforms of the elements of the matrix $\boldsymbol{\hat \chi}(\omega)$, decrease exponentially fast with time. In other words, they are {\it bona fide} causal  response functions and  $\det \boldsymbol{\hat \chi}(\omega)$ is thus analytic in the upper half of the complex $\omega$-plane.  From Eq. (\ref{Eqmu1}), one can readily show by using contour integration and  simple algebraic manipulations \cite{SM} that $\mu(\beta_L)=\gamma_L/m$, in agreement with the IFT. Hence, $g_Q(\beta_L)\equiv\lim_{t\to \infty}Z_Q(\beta_L,t)\exp[-\mu(\beta_L)t]=1$. The LDF for ${\cal Q}$ is thus simply given by the Legendre transform of $\mu(\lambda)$ and $I(q)\sim-\lambda_+ q$ for $q\to -\infty$, where $\lambda_+$ is the right boundary of the domain of definition of $\mu(\lambda)$ \cite{T2009}. 

When the hat dynamics does not converge, the corresponding response functions of the $N$ oscillators do not decrease with time (implying for instance that the variances of $u_i$ and $v_i$ grow indefinitely with time). In this case, the elements of the matrix $\boldsymbol{\hat \chi}(\omega)$ do {\it not} correspond to the Fourier transform of standard response functions. (We recall that $\boldsymbol{\hat \chi}(\omega)$ is just obtained by changing $\gamma_L$ into $-\gamma_L$ in $\boldsymbol{\chi}(\omega)$.) This shows up in the fact that $\det \boldsymbol{\hat \chi}(\omega)$ has also poles in the upper half complex plane. Then, $\mu(\beta_L)< \gamma_L/m$ as can be seen in Fig. 1. This implies that the actual SCGF $\mu_Q(\lambda)$ differs from $\mu(\lambda)$ for this special value of $\lambda$ and is therefore discontinuous in $\beta_L$. (On the other hand, in spite of the symmetry $\mu(\lambda)=\mu (\beta_L-\beta_R -\lambda)$,  $\mu_Q$ is continuous in $-\beta_R$, with $\mu_Q(-\beta_R)= \mu(-\beta_R)= \mu(\beta_L)\ne \gamma_L/m$.) As a result, 
\begin{align}
Z_{Q}(\beta_L,t)e^{-\mu(\beta_L)t} \sim e^{[ \gamma_L/m-\mu(\beta_L)]t}  \to \infty
\end{align}
when $t\to \infty $, which implies that $g_Q(\lambda)$ diverges at $\lambda=\beta_L$. Then, the leading contribution to the LDF comes from the singularity in $\beta_L$ and $I_Q(q)=-\mu(\beta_L)-\beta_L q$ for $q\le -\mu'(\beta_L)$ \cite{noteFI}. Note that the existence of this singularity, which actually results from the average over the final degrees of freedom at time $t$, is predicted without investigating the analytical properties of  $g_Q(\lambda)$, whose expression is quite involved \cite{KSD2011} except for $N=1$ \cite{Sab2012}.  We also stress that the IFT tells nothing about the presence of one (or more) other pole(s) in $g_Q(\lambda)$ that is associated with the average over the initial state and whose location depends on the choice of $p({\bf x}^0)$. When this pole exists and belongs to the domain of definition of $\mu(\lambda)$, there is also an exponential tail in the right wing of $P_Q(Q)$ (see \cite{Sab2012,MR2012} for $N=1$).
 
We now  consider the fluctuations of the medium entropy production $\Sigma_m[{\bf X}]$ and of the total entropy production $\Sigma[{\bf X}]$. The latter satisfies the standard IFT obtained by time reversal, $\langle e^{-\Sigma[{\bf X}]}\rangle_{st}=1$, whereas the former follows the IFT derived above, $\langle e^{-\Sigma_m[{\bf X}]}\rangle_{st}=e^{\frac{\gamma_L+\gamma_R}{m}t}$. The corresponding auxiliary dynamics amounts to reversing the sign of both $\gamma_L$ and $\gamma_R$, which implies that this dynamics never leads to a steady state \cite{SM}. In addition, the long-time behavior of the characteristic functions $Z_{\Sigma_m}(\tilde\lambda,t)\equiv\langle e^{-\tilde\lambda\Sigma_m}\rangle_{st}$ and $Z_{\Sigma}(\tilde\lambda,t)\equiv \langle e^{-\tilde\lambda\Sigma}\rangle_{st}$ is given by formulas similar to Eq. (\ref{EqZA}), but with $\mu(\lambda)$ replaced by $\mu(\tilde\lambda[\beta_L-\beta_R])$ \cite{SM}. From the expression of $\mu(\lambda)$ given above, one then readily finds for $\tilde\lambda=1$ that $\mu(\beta_L-\beta_R)=0$. As a result, $g_\Sigma(1)=1$  but $g_{\Sigma_m}(1)$ diverges, which leads to  an exponential tail in the corresponding pdf, with $I_{\Sigma_m}(\sigma_m)=-\sigma_m$ for $\sigma_m\le -\langle \sigma \rangle_{st}$ where $\langle \sigma \rangle_{st}=(\beta_L-\beta_R)^2T_LT_R\int d\omega\: {\cal T}(\omega)/(4\pi)$ is the average entropy production rate.

\subsection{Non-Markovian feedback control}

As a second example, we consider a non-Markovian dynamics governed by the time-delayed Langevin equation \cite{MR2014,RMT2015}
\begin{align}
\label{EqLD}
m\dot v_t=-kx_t+k'x_{t-\tau}-\gamma v_t+\xi_t \,,
\end{align}
with $\langle \xi(t)\xi(t')\rangle=2\gamma T\delta (t-t')$, which describes the motion of feedback-cooled nano-mechanical resonators in the vicinity of their fundamental mode resonance ({\it e.g.}, the cantilever of an AFM \cite{Mont2012}). Due to the delay $\tau$, the dissipated heat ${\cal Q}[{\bf X},{\bf Y}]=\int_0^t dt' \: [\gamma v_{t'}-\xi_{t'}]v_{t'}=-\int_0^t dt'\: [m\dot v_{t'}+kx_{t'}-k'x_{t'-\tau}]v_{t'}$ and the work done by the feedback force ${\cal W}[{\bf X},{\bf Y}]=k'\int_0^t dt' \: x_{t'-\tau} v_{t'}={\cal Q}[{\bf X},{\bf Y}]+\Delta {\cal U}({\bf x}^t,{\bf x}^0)$  also depend on the trajectory in the time interval $[-\tau,0]$, which is denoted  by ${\bf Y}$. We also define  a  ``pseudo" entropy production \cite{RTM2015b} $\Sigma[{\bf X},{\bf Y}]=\beta{\cal Q}[{\bf X},{\bf Y}] + \ln [p_{st}({\bf x}^0)/p_{st}({\bf x}^t)]$ where $\beta=1/T$. 

By using again Fourier transforms, we obtain  \cite{SM}
\begin{align}
\label{Eqmuw1}
\mu(\lambda)= -\frac{1}{4\pi} \int_{-\infty}^{\infty} \frac{d\omega}{2\pi} \ln[1-4\lambda k'\gamma T \omega \sin(\omega \tau)\vert\chi(\omega)\vert^2] \,,
\end{align}
where $\chi(\omega)=[-m\omega^2-i\gamma\omega+k-k'e^{i\omega\tau}]^{-1}$ is the Fourier transform of the response function of the time-delayed oscillator.  The expression of $\mu(\beta)$ is then given by Eq. (\ref{Eqmu1}) with $\hat \chi(\omega)=\chi(\omega)\vert_{\gamma \to -\gamma}$, and we are again interested in the  behavior of the system under the auxiliary hat dynamics to predict the possible existence of a singularity in the characteristic functions $Z_Q(\lambda,t),Z_W(\lambda,t)$ in $\lambda=\beta$ and $Z_{\Sigma}(\tilde\lambda,t)$ in $\tilde\lambda=1$.
\begin{figure}[hbt]
\begin{center}
\includegraphics[width=8cm]{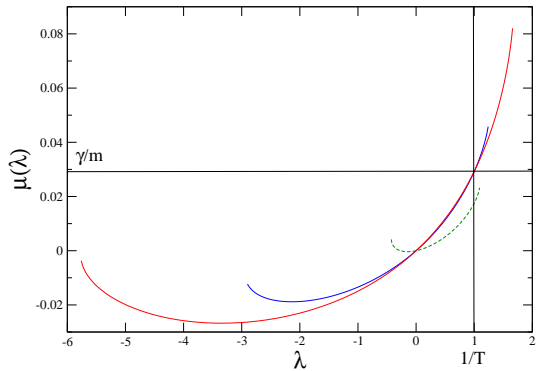}
 \caption{\label{Figdelay} (Color on line) $\mu(\lambda)$ for the time-delayed Langevin equation (\ref{EqLD}) [$\mu(\lambda)$ is defined in $(\lambda_-,\lambda_+)$ with diverging slopes at the boundaries]. The oscillator quality factor is $Q_0=\sqrt{mk}/\gamma=34.2$ \cite{GBPC2010} and $k'/k=0.25$. Each curve corresponds to a different value of the delay: $\tau =7.6$ (blue), $7.9$ (red), $8.4$ (green dashed). For $7.37\le  \tau  \le 8.32$, $\mu(\beta)=\gamma/m=1/34.2$.}
\end{center}
\end{figure}

Fig. 2 shows the influence of the delay $\tau$ on $\mu(\lambda)$  for a feedback-cooled resonator operating in its second stability lobe (see \cite{RMT2015,RTM2015b} for details).  The quality factor $Q_0$  corresponds to the AFM micro-cantilever used in the experiments of \cite{GBPC2010}.  
It is found that a stationary regime is reached with the hat dynamics for  $7.37\le  \tau  \le 8.32$ (with $\omega_0^{-1}=\sqrt{m/k}$ taken as the time unit). Then, $\mu(\beta) =\gamma/m$ as a result of the causal character  of $\hat \chi(\omega)$. Hence, $g_Q(\beta)=\lim_{t\to \infty}Z_Q(\beta,t)\exp[-\mu(\beta)t]=1$. Moreover, by inserting Eq. (\ref{Eqratio}) (or rather its extension to the case with delay \cite{SM}) in the expression of the characteristic function of the work, one derives that 
\begin{align}
\label{EqgW}
g_W(\beta)&=\lim_{t\to \infty}Z_W(\beta,t)e^{-\mu(\beta)t}\\
&= \int d {\bf x}^0 \:p_{st}({\bf x}^0)\int d{\bf x}^t\: e^{-\beta \Delta {\cal U}({\bf x}^t,{\bf x}^0)}\hat p_{st}({\bf x}^t)\,, \nonumber
\end{align}
which is numerically found to be finite. The LDF's for ${\cal Q}$ and ${\cal W}$ are thus simply given by the Legendre transform of $\mu(\lambda)$ and the  left wings of $P_Q(Q)$ and $P_W(W)$ are asymptotically identical. On the other hand,  \begin{align}
Z_{\Sigma}(\beta,t)e^{-\mu(\beta)t}\to \int d {\bf x}^0 \int d{\bf x}^t\: p_{st}({\bf x}^t)\hat p_{st}({\bf x}^t)
\end{align}
as $t\to \infty$, which is clearly diverging. This leads to an exponential tail in the pdf of $\Sigma$ with $I_\Sigma(\sigma)=- \gamma/m-\sigma$ for $\sigma\le-\beta \mu'(\beta)$.

When the hat dynamics does not converge, one has $\mu(\beta)< \gamma/m$ as before (see Fig. \ref{Figdelay}). This implies that $g_Q(\lambda)$ diverges at $\lambda=\beta$, which leads to an exponential tail in the LDF with $I_Q(q)=-\mu(\beta)-\beta q$ for $q\le -\mu'(\beta)$. (On the other hand, we cannot conclude for $g_W(\beta)$ and $g_{\Sigma}(1)$.) Note that $\beta$ is {\it not}  generically a pole of $Z_{\Delta {\cal U}}(\lambda,t)$, the characteristic function of $\Delta {\cal U}$ \cite{SM}. In other words, ${\cal Q}={\cal W}-\Delta {\cal U}$ cannot be treated as the sum of two independent random variables for very large $t$, with the poles attributed to $\Delta {\cal U}$, as is often done \cite{BJMS2006,TC2007,JGC2007, GSPT2013}.

\section{Concluding remarks} These two examples illustrate the usefulness of the new IFT for predicting the occurrence of extreme events  associated with exponential tails in the pdfÕs for the heat and related quantities.  

However, the above calculations were restricted to the case of linear systems for which one can solve the equations of motion by Fourier transform and study the analytic behavior of  the response function(s). There is a need for a more general  argument which we now offer. Let us take again the example of a one-dimensional chain connected to two heat reservoirs, but now with arbitrary interactions. Assuming that the system has reached a NESS, the characteristic function for the heat is given by
\begin{align}
&Z_Q(\lambda,t)=\int d{\bf x}^0\: p_{st}({\bf x}^0)\int d{\bf x}^t\int_{{\bf x}^0}^{{\bf x}^t} {\cal D} {\bf X}\:e^{-\lambda{\cal Q}[{\bf X}]}{\cal P}[{\bf X}\vert {\bf x}^0] \nonumber\\
&=e^{\frac{\gamma_L}{m}t}\int d{\bf x}^0 \: p_{st}({\bf x}^0) \int d{\bf x}^t \int_{{\bf x}^0}^{{\bf x}^t} {\cal D} {\bf X}\:e^{-(\lambda-\beta_L){\cal Q}[{\bf X}]}\hat {\cal P}[{\bf X}\vert {\bf x}^0] \ ,
\end{align}
where we have used Eq. (\ref{Eqratio}) to go from the first line to the second one. Differentiating with respect to $\lambda$  and using the IFT $Z_Q(\beta_L,t)=e^{\frac{\gamma_L}{m}t}$, we then obtain
\begin{align}
\label{Eqsumrule}
-\frac{\partial \ln Z_Q(\lambda,t)}{\partial \lambda}\vert_{\lambda=\beta_L}&=\int d{\bf x}^0 \:p_{st}({\bf x}^0)\int d{\bf x}^t  \nonumber\\
&\times\int_{{\bf x}^0}^{{\bf x}^t}{\cal D} {\bf X}\:{\cal Q}[{\bf X}]\hat{\cal P}[{\bf X}\vert {\bf x}^0] \ .
\end{align}
The r.h.s. of this equation is the average of the stochastic heat  ${\cal Q}[{\bf X}]\equiv {\cal Q}_L[{\bf X}]=\int_0^t dt'\: [-m\dot v_1+F(u_1,u_2)]v_1$ (see Eq. (\ref{EqQ})), but the dynamics of the first particle is now governed by the conjugate Langevin equation $m\dot v_1=F(u_1,u_2)+\gamma_L v_1+\xi_1$ where $\gamma_L$ has been changed to $-\gamma_L$. This leads to 
the exact relation
\begin{align}
\label{Eqheat5}
\frac{\partial \ln Z_Q(\lambda,t)}{\partial \lambda}\vert_{\lambda=\beta_L}&=\frac{\gamma_L}{m}\int_0^t dt'\: (\hat T^{(v)}_1(t')+T_L)
\end{align}
where $\hat T^{(v)}_1(t)\equiv m \langle \hat v_1^2\rangle$ denotes the instantaneous (kinetic) temperature of the first particle under the conjugate dynamics.  Since $Z_Q(\lambda,t)\sim e^{\mu_Q(\lambda)t}$ in the long-time limit, Eq. (\ref{Eqheat5}) implies that 
\begin{align}
\mu_Q'(\lambda=\beta_L)=\frac{\gamma_L}{m}\lim_{t \to \infty}\frac{1}{t}\int_0^t dt'\: (\hat T^{(v)}_1(t')+T_L)\ .
\end{align}
Now, again, two cases may occur. First, the hat dynamics converges to a stationary state, which means that $\hat T^{(v)}_1(t) \rightarrow \hat T^{(v),st}_1$ (typically exponentially fast). Then $\mu_Q'(\lambda=\beta_L)=(\gamma_L/m)(\hat T^{(v),st}_1+T_L)$, which is a finite quantity. Second, the  hat dynamics does not converge, and $\hat T^{(v)}_1(t)$ increases indefinitely with time. Then $\mu_Q'(\lambda=\beta_L)$ diverges, which indicates that $\mu_Q(\lambda)$ is discontinuous in $\lambda=\beta_L$. This is the typical signature of rare events in which a large amount of heat is flowing to the system, leading to an exponential tail in the left ({\it i.e.}, negative) wing of the pdf.

In conclusion, we have shown that, for a given model,  one can avoid the difficult detection of rare events by checking instead whether the conjugate ``hat"  dynamics leads to a stationary regime or not, a much  easier numerical task. Finally, we also note that the calculations for the harmonic chain suggest that such tails may be absent in the case of sufficiently strongly interacting particles, which would be interesting to check experimentally.

\begin{acknowledgments}
M.L. R. is grateful to A. Dhar and K. Saito for stimulating discussions during the workshop ``New Frontiers in Non-equilibrium Physics 2015" held at the Yukawa Institute for Theoretical Physics in Kyoto.
\end{acknowledgments}

\end{document}